\documentclass[article]{elsarticle}

\usepackage{lineno,hyperref}
\modulolinenumbers[5]

\usepackage{enumerate}
\usepackage{tikz}

\usepackage{svg}
\usepackage{gensymb} %degree symbol

\journal{Journal}

\usepackage{amsfonts}

\usepackage{amsmath}
\usepackage{appendix}
\usepackage{graphicx}% Include figure files
\usepackage{dcolumn}% Align table columns on decimal point
\usepackage{bm}% bold math
\usepackage{enumitem}
\usepackage{natbib}
\usepackage{xcolor}
\usepackage[
top=1in, bottom=1in, left=1.5in, right=1.5in,
]{geometry}

\usepackage{subfig}
\usepackage[rawfloats=true]{floatrow}

\newcommand{\vince}[1]{\textcolor{black}{#1}}

\usepackage{soul}

\begin{document}

	\begin{frontmatter}
			
\title{
Hierarchical physically based machine learning in material science: the case study of spider silk}

\author{Vincenzo Fazio$^1$,  Nicola Maria Pugno$^{1,2*}$,\\ Orazio Giustolisi$^{3}$, Giuseppe Puglisi$^{3**}$.}%%%% Author details

\address{$^{1}$ Laboratory for Bioinspired, Bionic, Nano, Meta Materials \& Mechanics,\\ University of Trento, Via Mesiano 77, 38123 Trento, Italy;}
\address{$^{2}$ %\\ Ket Lab, Edoardo Amaldi Foundation, Italian Space Agency, Via del Politecnico snc, 00133 Rome, Italy;
	School of Engineering and Materials Science, Queen Mary University of London,\\ Mile End Road, London E1 4NS, U.K.;}
\address{$^{3}$ Department of Civil Environmental Land Building Engineering and Chemistry, Polytechnic University of Bari, via Orabona 4, 70125 Bari, Italy.}

\address{$^{*}${nicola.pugno@unitn.it}}
\address{$^{**}${giuseppe.puglisi@poliba.it}}

\begin{abstract}

Multiscale phenomena are characterized by complex structure-function relationships emerging where entities at different scales aggregate into structures with unexpected final properties.
Mathematical modelling for multiscale phenomena typically requires the deduction of a set of differential equations at different scales that predict the macroscopic behavior. The complexity of these equations and the number of essential parameters make developing effective, predictive models challenging. To address this challenge, recent scientific literature has explored the possibility of taking advantage of the availability of sophisticated numerical techniques in the artificial intelligence and machine learning fields.

Here, we focus on a fundamental aspect in multiscale phenomena, {\it i.e} the recognition of the hierarchical role of variables. In the spirit of the considered numerical approaches and especially in determining efficient tools to deduce effective analytical relations for material modelling, we focus, in a Pareto front interpretation, on the determination of simple accurate relations, starting from experimental multiscale analyses. From a physical point of view, the aim is to deduce information at higher scales from lower scales data, possibly respecting their hierarchical order. A crucial aspect of the proposed approach is the deduction of causality relations among the different variables to be compared with the available theoretical notions and possibly new interpretations resulting by the data modelling. This result in a stepwise approximation going from data modelling to theoretical equations and back to data modelling.

To demonstrate the key advantages of our multiscale numerical approach, compared to classical, non-physically based data modelling techniques, we consider the explicit example of spider silk --a material with exceptional properties hugely analyzed also in the spirit of bioinspiration and strictly based on an evolutionary hierarchical optimization--. The description of the complex behavior of such material and the importance of the structure organization at different scales represent an open problem in material science and within the field of the design of new bioinspired materials.
The remarkable macroscopic spider silk behavior is indeed the result of interesting mesostructures arising from the aggregation of amino acids at the molecular scale. 
The comparison with recent data modelling results, neglecting causality and the multiscale character of the examined material,  demonstrates the importance of the search for new data modelling approaches aiming at a growth of a more deep scientific knowledge in the field. 

We argue that, due to the generality of our results, our approach may represent a {\it proof of concept} in many fields where multiscale, hierarchical differential equations regulate the observed phenomenon.

\end{abstract}

\end{frontmatter}

	\section{Introduction}

%importanza modellazione multiscala. Disponibili nuove tecniche per investigare nano-scala
Multiscale models play a crucial role in different fields of theoretical and applied science, especially due to the increasing possibility of experimental analyses and technologies working down to the nanoscale such as Atomic Force Microscopy (AFM), optical tweezers, magnetic tweezers \cite{bustamante2000grabbing}, etc. The technological impact has been incredible, with the possibility of designing systems at very low scale NEMS, MEMS \cite{bhushan2007nanotribology,zhu2019development} also with the production of new smart systems based on a hierarchical organization at different scales: multiscale metamaterials \cite{ChenPugno2013}, multiscale gecko inspired adhesive systems \cite{carlo2006biomimetic}, self-cleaning lotus inspired surfaces \cite{solga2007dream,bae201425th}, etc.
As a matter of fact, in different fields a huge experimental literature delivering big data libraries on hierarchical systems, starting from the nano and micro scales, up to the macro scale, is now available. The scientific impact of such experimental results in new numerical/theoretical tools delivering correct physical interpretation in several fields of impact in engineering \cite{ChenPugno2013}, medicine \cite{ashley2016towards}, physiology \cite{zhang2012cross}, biology \cite{mclennan2012multiscale} and physics \cite{ji2022magnetic}.

% increasing attention in different fields, from complex systems such as aircraft to the field of materials, where it is allowing great advances in the design of new materials and meta materials and the full understanding of the behavior of intricate natural systems.
%%esempi da chen pugno 2013
%As examples, the gecko adhesion and spider silks, despite being studied for several years, have been effectively described only within the framework of the multiscale modelling \cite{ChenPugno2013,fazio2022spider}.

% introduzione tecniche digitali
Indeed, the analysis of the now available huge mole of experimental data can lead to a corresponding increase in the theoretical understanding and modelling of the resulting physical system, only if adequate numerical instruments of data modelling are available. We live in the digital age and the possibility of new instruments such as unprecedented power of calculation and machine learning techniques opened up the possibility of new incredible tools of analysis of these multiscale data. On the other hand, as in every transition, the digital transition brings significant risks and drawbacks if not deeply analyzed in its possible effects. Thus, machine learning can lead to a scientific knowledge growth or obscuration, rationalization or unclearness, access to deeper theoretical models or reliance on purely data mining approaches. 

In this work, we trace a rational way in the direction of deducing new tools for the modelling of multiscale phenomena based on machine learning techniques that may lead to real advances in scientific knowledge. Among many different techniques with the potentiality of significant growth in this perspective, the proposed approach started from the following observations on the state of the art in the field.
First, we point out that among data-driven techniques, many of which have been developed in recent years, Artificial Neural Networks (ANN) and Genetic Programming (GP) are the most adopted to model complex, non-linear processes including multiscale hierarchical phenomena.
Loosely speaking, ANN uses models consisting of multiple processing elements (neurons) connected by links of variable weights (parameters) to deduce typically `black-box' representations of the analyzed systems.
Learning in ANN involves adjusting the parameters (weights) of interconnections in a highly parametrized system. 

In few words, the main widely recognized disadvantages of ANN model construction are the curse of dimensionality, overfitting issues and parameter estimation \cite{haykin1999neural,giustolisi2005improving}. 
 The well-known curse of dimensionality refers to the exponential increase in the need of parameters when the model input space grows. This means that the number of connections exponentially raises and in a such widened space the training set of input becomes more sparse or the amount of data needed to preserve a constant level of accuracy increases exponentially. On the other hand, in such a way, ANN acquires greater flexibility in mapping events with complex structure. However, this leads often to overfitting problems, that is ANN tends to fit training data too precisely due to the large number of parameters resulting in the propensity to generate poor predictions for events not close to the training data set.
Another disadvantage of using ANN is the difficulty of incorporating knowledge derived from known physical laws into the learning process.

Despite these drawbacks, several significant results in this field have been reported. Here we recall the very interesting recent fundamental result of the use of machine learning techniques to predict protein topological conformations from amino acid sequences with high accuracy \cite{baek2021accurate}.
%cosa è stato fatto fino ad ora su AI-Multiscale phenomena
%The application of artificial intelligence for multiscale problems has already been considered in different scientific fields in the last years. 
In the field of multiscale materials modelling, we may recall that Gu et al. \cite{gu2018bioinspired} have employed finite elements analysis together with convolutional neural network algorithms to predict and optimize the toughness of hierarchical composite systems and validated their results through additive manufacturing and testing. 
\vince{Recently in \cite{linka2023new} the authors adopted ANN to chose among a class of specific constitutive models depending on the right Cauchy Green deformation tensor invariants, the model that best reproduces stress-strain behaviors under different classes of deformation. While the approach is interesting, it is highly oriented by the specific knowledge of the problem and restricted to the special case when the class of constitutive laws is already known: i.e. the stress dependence on the deformation invariants. }
As further machine learning application on materials science, a k-means clustering approach was employed to predict the behavior of heterogeneous materials under irreversible processes like inelastic deformations obtaining a data-driven, two-scale model \cite{liu2016self}. Also, Neural Network approaches were employed within the field of the meta-materials to perform tasks like topological optimization \cite{kollmann2020deep}, while Bayesian machine learning was employed in a data-driven design of metamaterial building blocks based on several design variables \cite{bessa2019bayesian}. 
Good predictive performances were obtained also by neural network methods in linking the elastic properties of composite materials to their mesoscale structure, in particular, the three-dimensional microstructure to its effective (homogenized) properties \cite{cecen2018material}. For steels, machine learning based microstructural analysis, property prediction, and properties-to-microstructure inverse analysis were conducted \cite{wang2019property}. 

The implications are thus fundamental and let us obtain relevant information for problems that have longly been theoretically unresolved, such as the recalled long-lasting problem of predicting the protein structures from amino acid sequences \cite{baek2021accurate}. On the other hand, the main drawback in the perspective of extending the knowledge for the theoretical modelling of such phenomena is that ANN leads to ``black-box" approaches. There is then a strong limitation on the `operational' advantages due to the lack of interpretability of the artificial intelligence results. Some very recent works address this issue \cite{murdoch2019definitions,du2019techniques}, but this is still an open problem \cite{doshi2017towards,molnar2020interpretable} due to the intrinsic nature of the approach, summarized above. 

On the other hand, GP is an evolutionary computing method that can generate a more `transparent' representation of the system. In particular, in this field, the symbolic regression is a technique proposed by Koza \cite{koza1992genetic} that creates mathematical expressions to fit a set of data points using the evolutionary process of the GP. 
In brief, like all evolutionary computing techniques, symbolic regression manipulates populations of solutions (in this case mathematical expressions) using operations analogous to the evolutionary processes observed in nature. The genetic programming procedure mimics natural selection through successive generations of solutions improving the fitting of the data points.
As a result, GP allows global exploration of expressions providing insights into the explicit relations between input and output data.

Here we explore the potential synergies between machine learning and multiscale modelling to produce robust predictive models that take into account the underlying physics to handle challenging problems in this field. 
In particular, the main purpose is to show that the multiscale character typically corresponds to a hierarchical organization and such hierarchy should be crucial in the process of data mining and data modelling, based on the underlying physical phenomena.
The present article, therefore, contributes to the definition of a physically based data modelling where the expression ``physically based'' here wants to highlight that although machine learning has been effectively used to automate data processing and ensure high accuracy and repeatability of outcomes of many physical phenomena, the most diffused techniques typically ignore basic physical principles, which may result in non effective approaches possibly leading to unphysical results. 

%intro e descrizione EPR
%Un paio di pagine, da Giustolisi 22, con riferimenti a biblio 2006 e 2009.
More in detail, here we consider the application of a well known symbolic data-modelling method named Evolutionary Polynomial Regression (EPR) \cite{giustolisi2006symbolic,giustolisi2009advances}, which enables the discovery of explicit and generalizable equations for the underlying physical model. 
Indeed, EPR could be defined as a ``gray box'' approach that, unlike typical ANN methods, is structured in such a way that it is possible to integrate information arising from data modelling with established physical knowledge of the phenomena. 
Specifically, an explicit advantage is the possibility of deducing the independent variables of the model at each scale and substantial properties of the functional dependencies among the involved variables. Interestingly, as we explicitly show in this paper, the use of EPR let us obtain a number of different analytical models, optimized using a multi-objective method. The choice among these models is then based on their physical interpretation and the effectiveness of the model. The new obtained relations can then be used as input for the successive numerical analysis, thus allowing for a continuous interchange between physical interpretation and data modelling.
Indeed, EPR generates symbolic and explicit nonlinear equations that include a small number of polynomial structural parameters.  \\

%CASE STUDY

To analyze the efficiency of the proposed approach in treating complex multiscale hierarchical phenomena, we here consider the field of constitutive modelling of complex material behaviors. Many biological examples of evolutionary material optimization represent the possibility of obtaining unreached material performances at the macroscale, based on a clever, hierarchical organization of weak composing materials at the lower scales \cite{keten2010nanoconfinement}. A further enrichment in biological structure is to possibly include different composing materials ~\cite{bosia2012investigating}.
The analytical description of how the macroscopic performances result from these complex low scale material organizations is far from being reached and represents an meaningful benchmark not only for their theoretical interest, but also in the crucial field of bioinspired material design \cite{arndt2022engineered,liu2016hierarchical}.

Specifically, we here consider the paradigmatic example of spider silk, one of the most studied natural materials due to its extreme mechanical properties, particularly its strength and toughness. The availability of increasingly sophisticated experimental techniques allowed for a deeper understanding of its complex multiscale, hierarchical material structure. Despite this, many relevant phenomena governing the strong material history dependence, rate, temperature, and humidity effects remain unknown. To be specific, our data modelling analysis is based on recent experimental observations on a large number of silks from different spider species from all over the world, where several material properties at different involved scales have been cataloged for the fist time in a comprehensive database\cite{arakawa20221000}. 

We refer to previous data modelling results that, while allowing qualitative interpretation of the results, cannot directly afford new analytic models. As a result, the availability of such new comprehensive experimental data represents for sure a fundamental possibility of gaining new insight in this multiscale model. This requires not only to refer to purely statistical properties of the available data, as in classical data modelling \cite{arakawa20221000}, but to adopt numerical techniques allowing for the deduction of analytical relations to be theoretically interpreted and implemented.

%FINE INTRODUZIONE
This work aims to be general within the framework of a multiscale description of physical phenomena and the deduction of larger scales properties from the structures at lower scales. Indeed in the formulation of the specific case study here analyzed, we have considered three scales starting from the {\it micro} (protein) scale, to the {\it macro} scale passing through the {\it meso} scale. We explicitly impose in our approach that these three scales interact with each other in a hierarchical way. In particular, we consider the three possibilities of deduction of the meso from the micro properties, a successive macro from meso and eventually an interesting direct micro to macro deduction.

This is then, in our opinion, a first step toward a more effective adoption of the new availability of data and data modelling techniques that can be of fundamental help in several fields of multiscale phenomena when dealing with a high number of experimental data.

\section{Evolutionary Polynomial Regression}
In this section we give, for the help of the reader, a brief introduction to the mathematical treatment of numerical optimization problems based on EPR algorithms. We refer to \cite{giustolisi2006symbolic,giustolisi2009advances} for a detailed description of the method.

EPR method generates explicit mathematical expressions to fit a set of data points starting from the symbolic equation\begin{equation}
{\bf Y}=\sum_{j=1}^m f(\mathbf{X},g(\mathbf{X}),a_j)+a_0 \label{eq:epr}
\end{equation}
where ${\bf Y}$ is the vector of output dependent variables, $\mathbf{X}$ is the vector of input variables,  $f$ is a polynomial function composed of $m$ terms generated by the algorithm, linearly depending on the unknown $a_j$ parameters, $g$ is a function defined by the user (in our case we will consider power laws with each input variable ${\bf X}_i$ raised to an exponent varying among an a priory fixed set of numbers), plus the bias term $a_o$. 

Thus in the case of power expressions considered in the following, EPR technique generates formulae of variable number of polynomial terms, performing a global search of the expression of symbolic expressions for $f$. Synthetically, EPR is performed in two steps: a) structure identification and b) parameter estimation. The first stage entails simultaneously determining the best `arrangement' of the independent variables and the related exponents. A multi-objective genetic algorithm termed OPTIMOGA, which stands for Optimized Multi-Objective Genetic Algorithm, is used to finalize this optimization. This algorithm has three targets: maximization of the so called fitness function (in other words a measure of how closely the regression expression fits the data points), minimization of the number of polynomial coefficients, and reduction of the number of inputs.
Observe that, since the user defines a priori the set of candidate exponents, the possible negligible input variables are obtained by including zero as a candidate exponent. This allows for the fundamental aspect, recalled in the introduction, of determining the effective independent variables. The values of the parameters $a_j$ are determined in a second stage using the linear Least Squares (LS) approach, which minimizes the Sum of Squared Errors (SSE). In addition to the usual LS search, the LS is performed by searching for only positive values (constrains $a_j>0$) to avoiding overfitting, by excluding sequences of terms with negative/positive $a_j$ values that may result from the modelling of the data noise \cite{giustolisi2007multi}. 
%\vince{In addition to the usual LS search, the LS is performed by searching for only positive values (constrains $a_j>0$), because negative terms usually have a poor physical meaning, as they simply balance positive terms returning a better description of noise \cite{giustolisi2007multi}.  \orazio{In other words, such constraint contribute to avoiding overfitting by excluding sequences of terms with negative/positive $a_j$ values that may result from the modelling of the data noise. On the other hand, this condition may be assumed as physical in the sense of incremental contribution to the phenomenon description through functional groups of decreasing importance.}
\vince{Moreover, the uncertainty of the coefficients ($a_j$) is evaluated  during the search and the distribution of estimated pseudo-polynomial coefficients is used to eliminate those parameters whose value is not sufficiently larger than zero \cite{giustolisi2004novel,giustolisi2006symbolic}.
Indeed, it may be argued that a low coefficient value with respect to the variance of estimates relates to terms that describe noise rather than the underlying function of the phenomenon being studied.
}

As a starting point the candidate independent variables, the general polynomial structure, functions, exponents, and the maximum number of terms are assigned based on the starting knowledge of the physical phenomenon. The exponents can reflect the types of relationships between the inputs and output. For example, if the vector of candidate exponents is chosen to be ${\bf EX}=[-1,-0.5,0,0.5,1]$, the maximum number of terms is $m=4$ and if the candidate independent input variables are $k=3$, the polynomial regression problem is to find a matrix of exponent ${\bf ES}_{4 \times 3}$. 
In a first stage, an initial population of matrix of exponents is generated.
An example of such a matrix is 
\begin{equation}
{\bf ES}_{4 \times 3}=\begin{bmatrix}
1 & 0.5 & 0 \\
0 & 0 & 1  \\
0 & -0.5 & 1  \\
-1 & 0 & 0.5  
\end{bmatrix}
\end{equation}
so that the expression \eqref{eq:epr} is given as:
\begin{equation}
{\bf Y}=a_o + a_1 \,{\bf X}_1 \, {\bf X}_2^{0.5} + a_2\, {\bf X}_3 +a_3\, {\bf X}_2^{-0.5} \, {\bf X}_3 + a_4\, {\bf X}_1^{-1} \, {\bf X}_3^{0.5}
\end{equation}
The adjustable parameters $a_j$ are then computed by minimizing the SSE as a cost function. 
It follows the evaluation of the fitness function: if the termination criterion is satisfied, the output results are shown, otherwise a new matrix of exponents is generated through Genetic Algorithm (GA) including crossover, mutation, ranking selection \cite{giustolisi2006symbolic} and again the adjustable parameters are calculated and the fitness function evaluated until the termination criterion is satisfied.

Interestingly, for a given data set of observations, a regression-based technique needs to search among an infinite number of possible models to explain those data.
Among different models that equally perform for the description of a given phenomenon the simplest one is chosen (Occam's razor or principle of parsimony). The approach then considers both accuracy and simplicity, that is it searches for an effective model describing the data as well as a simple expression {\it i. e.} with a small number of inputs and polynomial coefficients. \vince{The small number of constants to be
estimated helps to avoid overfitting problems,
especially for small datasets.}
The equally performing models are those composing the Pareto dominance front \cite{pareto1896cours,giustolisi2009advances} and since EPR returns the whole set of formulae of the Pareto front, the final choice of the model among different possible relations is then based on physical considerations \cite{giustolisi2004using}. 
\vince{The selection of a symbolic model through the reading of a Pareto front of similar but increasingly complex models brings into play the knowledge of the phenomenon and the insight of the user as opposed to a mere decision on the acceptance of the single model generated by classical statistical regression frameworks.}

In our opinion, this is a fundamental aspect because lets us understand the effective functional dependence among the different involved variables. This is why EPR can be categorized as a grey-box approach because it displays the link between inputs and outputs with an explicit expression, whose consistency can be easily analyzed and understood based on physical insight. More explicitly it allows the possibility of analyzing physical causality among the variables.
In this respect, we observe that GP generates formulae/models for $f$, coded in tree structures of variable size, performing a global search of the expression for $f$ as symbolic relationships among ${\bf X}$ while the parameters $a_j$  play a  role only in the optimization process. On the other hand, ANN goal is to map $f$, without focusing on the level of knowledge of the functional relationships among ${\bf X}$. This is why we argue that EPR represents a better tool for data-driven knowledge discovery. 

 %In this context, unlike to ANNs, the prior knowledge and fit technique of EPR generate fewer parameters, making it less susceptible to overfitting. 

\section{Case study: spider silk}

Spider silk is one of the most studied natural materials due to its extreme mechanical properties, particularly its strength and toughness, which outperform many high-performance man-made materials. Furthermore, spider silks are regarded as the foundation of a new class of high-performance fibers in the context of biomimetics~\cite{greco2021tyrosine,arndt2022engineered}.
The availability of increasingly sophisticated experimental techniques has allowed for a deeper understanding – both chemically and structurally – of the complex multiscale, hierarchical material structure at the heart of their notable mechanical behavior over the last few decades. Despite this, many relevant phenomena governing their loading history dependence, rate, temperature, and humidity effects remain unknown, particularly when multiscale effects are taken into account~\cite{review}.

At the molecular level, spider silks are made up of an amorphous matrix of oligopeptide chains and pseudo-crystalline regions composed primarily of polyalanine $\beta$-sheets \cite{elices_hidden_2011, sponner_composition_2007} with dimensions ranging from $1$ to $10$ nm \cite{keten_nanostructure_2010}, mostly oriented in the fiber direction \cite{jenkins_characterizing_2013}. The radial cross section of the fiber is highly organized \cite{li-new-1994,eisoldt-decoding-2011,sponner_composition_2007}. Furthermore, the chemical and structural composition varies according to the different silks produced by the different glands and, of course, the different species. Here, we focus on the most performing and extensively studied type of silk known as {\it dragline}.
%More in detail, the thread is covered by a skin~\cite{yazawa_role_2019}.

	%breve descrizione composizione alle diverse scale di spider silk.
	\subsection{Micro scale}
Spider dragline silk fibers (also known as Major Ampullate silk) are constituted by structural proteins called Spidroins, which are divided into two major subtypes, MaSp1 and MaSp2. The overall sequence architectures of the two subtypes are similar, with a highly repetitive core region flanked by small N-terminal and C-terminal domains (NTD and CTD, respectively). The repetitive regions, which account for 90$\%$ of the primary structure, are composed of alternating runs of polyalanine and multiple glycine-rich motifs arrayed in tandem. Moreover, very recent studies, prompted primarily by advances in proteomics and sequencing technologies, paint a more complex picture of dragline silk composition than a simple MaSp1/MaSp2 dichotomy \cite{arakawa20221000}. However, despite the complexity of the composition of the spider silks, here we only consider the protein MaSp1 and MaSp2 which are widely recognized as the two main composing the spider silk.
From the secondary structure point of view, the MaSp1 is mainly organized into pseudo-crystalline polyalanine $\beta$-pleated sheets \cite{li-new-1994,brown_critical_2011}. On the other hand, the MaSp2 is mainly constituted by proteins with a proline content preventing the formation of $\beta$-sheet crystals \cite{sponner_composition_2007} resulting in a structure with significantly lower crystallinity and macromolecules with weaker crystal domains, typically in the form of $\alpha$-helix and $\beta$-turns \cite{sponner_composition_2007,nova_molecular_2010}. 

We remark that as recognized in polymer mechanics \cite{flory_1982} and described also for the spider silk case in \cite{fazio2022spider}, the number of monomers of the macromolecule (i.e. protein for the silk case) is fundamental for the mechanical behavior of the material. Based on the fact that (i) the mechanical behavior of the spider silk material is to be ascribed to the repetitive region features more than the terminal region of the protein \cite{hayashi1999hypotheses}, and (ii) the pseudo-crystalline $\beta$-sheets, mainly present in the MaSp1, are recognized to be the most impactful feature in determining the exceptional strength of the spider silk \cite{yarger2018uncovering}, here we consider the following three quantities describing the protein scale of the silk material:
\begin{itemize}
	\item length of the repetitive region of the protein MaSp1 in terms of number of amino acids
	\item length of the repetitive region of the protein MaSp2 in terms of number of amino acids
	\item length of the polyalanine $\beta$-sheet in the protein MaSp1 in terms of number of alanine amino acid
\end{itemize}

	\subsection{Meso scale}
At the meso scale we consider the proteins' secondary structure, how macromolecules are arranged in the fiber and properties regarding the chemical and structural stability of the polymer.
In particular, we analyze the following material properties:
	\begin{itemize}
		\item Birefringence. It reflects the degree of molecular orientation of silk protein chains. The birefringence of the dragline silk fiber was calculated from the retardation value and silk fiber diameter \cite{arakawa20221000}.
		\item Degree of crystallinity. It was calculated based on wide-angle x-ray scattering (WAXS) analysis \cite{arakawa20221000}. In particular, it was obtained as the ratio of the total area of the separated crystalline scattering components to that of the crystalline and amorphous scattering components as resulting from the 1D profile obtained by the two-dimensional (2D) diffraction.
			\item Degradation temperatures. This quantity gives a measure in the chemical and structural stability of the silk. In \cite{arakawa20221000} the thermal degradation temperature has been defined as the temperature that yielded 1\% weight losses in the silk samples. Indeed, heating leads to changes of the molecular weight that in turn decreases the mass due to the production of gaseous byproducts of the chemical reactions. 
%At high temperatures, the components of the long chain backbone of the polymer can break (chain scission) and react with one another (cross-link) to change the properties of the polymer. These reactions result in changes to the molecular weight (and molecular weight distribution) of the polymer and can affect its properties by causing reduced ductility and increased embrittlement, chalking, scorch, colour changes, cracking and general reduction in most other desirable physical properties.[wiki]
	\end{itemize}	

%The water content was calculated from the percent weight loss associated with the evaporation of bound water from the TGA data based on a previous silkworm silk study

		\subsection{Macro scale}
Spider silk is a very interesting material from the point of view of its mechanical performance at the macroscopic scale. In particular, here we focus on the material stiffness and strength. The stiffness, of the order of tens of GPa, is above man-made polymers and at the top among other natural materials. The strength is even more interesting, being comparable with high strength steels (order of magnitude of 1 GPa)  and with the most performing man-made composites like the carbon and kevlar reinforced composites \cite{ashby2013materials}. %[citare asbhy, interessanti le figure da far vedere a Geppe ma non da inserire nell'articolo]
The reason for these so outstanding properties with respect to standard materials, is not yet clear, with an relevant role also in the extremely small diameter of dragline spider silk \cite{porter2013spider}. For this reason, we also consider the diameter in the properties at the macro scale. Finally, we address the very significant role of hydration in the material behavior of spider silks. Indeed, a striking effect observed in spider silks is the so called {\it supercontraction effect}, addressed, to the knowledge of the author, for the first time in 1977~\cite{Work_1977}, that occurs when a spider silk thread is exposed to humidity. Depending on the silk composition, the experiments show the existence of a Relative Humidity (RH) threshold beyond which the fiber contracts up to half of its initial (dry) length. This also results in the possibility of exploiting the supercontraction in the actuation field \cite{fazio2023water}. 
The experimentally observed contraction depends on several factors, including spider species~\cite{boutry2010evolution}, type of silk (among the up seven different ones that some spiders can produce \cite{vollrath1992spider,gosline1994elastomeric}), environmental conditions \cite{plaza2006thermo} and hydration rate \cite{agnarsson2009supercontraction}.
 The quantities we consider at the macro scale are therefore the following:
\begin{itemize}
\item Young's modulus, obtained from the stress-strain curves determined through tensile tests of single dragline silk fibers conducted at 25$^\circ$C and RH$\approx 50\%$ \cite{arakawa20221000}.
\item Tensile strength, calculated as the breaking force determined by tensile test divided by the undeformed cross-sectional areas of the fiber samples determined by SEM observations \cite{arakawa20221000}.
\item Diameter, determined by SEM observations \cite{arakawa20221000}.
\item Maximum supercontraction, calculated as  $(L_0 – L_f)/L_0$, where $L_0$ is  the length in dry  condition and $L_f$ in fully wet conditions (RH=100\%)  \cite{arakawa20221000}.
\end{itemize}

\section{Modelling strategy}	

For all the EPR run the maximum number of terms was set to $3$ to avoid overfitting and allow the aggregation of the inputs. The chosen set of candidate exponents is $[-1, -0.5, 0, 0.5, 1]$, adopted to keep the expressions simple.
%Regarding the configuration of the regression, preliminary tests performed with several structures revealed that the performance were not meaningfully affected. The simple formula reported in the Table were therefore chosen. 
Moreover, the expressions were optimized with a bias term $a_o$ since this element may compensate for the possible lack of relevant inputs in the model.

As training data, we employed the experimental results recently published in \cite{arakawa20221000}\footnote{Each quantity is considered with the unit of measurement reported in the original database, namely GPa for Young's modulus and limit stress, $\mu m$ for the diameter, $^\circ$C for the thermal degradation temperature, and number of amino acids for all the micro scale properties, whereas the supercontraction and the crystallinity are two nondimensionional quantities ranging in (0,1).
We remark that the information regarding the dimension of each term of the expression found by the EPR algorithm are included within the parameters $a_j$ (see Eqn. \eqref{eq:epr}) estimated by means the minimization of the SSE.}.
The above described quantities are summarized in  Table~\ref{tab:quantities} with the corresponding adopted symbol (the type of font distinguishes the scales):

\begin{table}[h]\centering
\begin{tabular}{lcl}
Micro scale	&	$a$	&	Length of the repetitive region of MaSp1 	\\
 					&	$b$	&	Length of the repetitive region of MaSp2	\\
					&	$c$	&	Length of the polyalanine $\beta$-sheet in the MaSp1 \\
\hline
Meso scale	&	$A$	&	Crystallinity	\\
 					&	$B$	&	Birefringence	\\
					&	$C$	&	Thermal degradation temperature (1\% loss)\\
\hline
Macro scale	&	$\mathbb{A}$	&	Young's modulus	\\
					&	$\mathbb{B}$	&	Tensile strength	\\
					&	$\mathbb{C}$	&	Diameter				\\
					&	$\mathbb{D}$	&	Supercontraction	\\
\end{tabular}
\caption{\label{tab:quantities} Material properties considered for the data modelling case study divided by scales.}\end{table}

\begin{figure}	\centering
	\includegraphics[width=.5\textwidth]{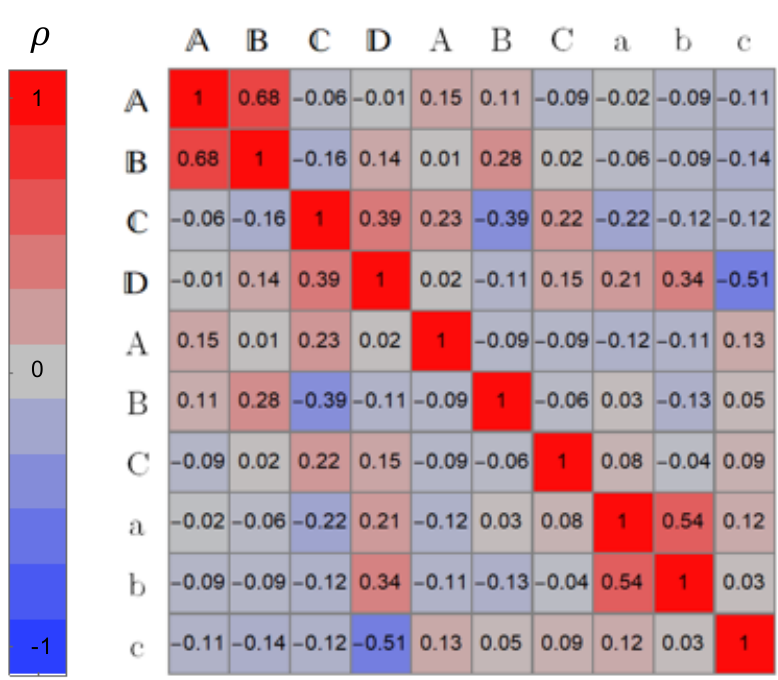}
	\caption{\label{fig:MatCorrExp} Experimental Pearson correlations among the material properties at the three scales considering all the silk reported in \cite{arakawa20221000}.}
\end{figure}

As a main parameter of accuracy, we report for the different numerical results the Coefficient of Determination\footnote{Here we consider the classical definition $R^2=1-\sum_{i=1}^N\frac{(x_i^{num}-x_i^{exp})^2}{(x_i^{exp}-\bar x^{exp})^2}$, where the $x_i^{num}$ are the output variables of the numerical test and $x_i^{exp}$ are the corresponding experimental values, with $i=1,...N$, where $N$ is the number of experimental observations considered as dependent variables.}. On the other hand, 
EPR also considers other, not explicitly reported here, indicators of performance, {\it e.g.} the sum of squared errors (SSE). As a result
$R^2$ does not necessarily increase as the complexity of the expressions grows.
%inserire qui considerazioni sugli indici delle performance
%riprendere da qui
The physical valence of the expressions found is discussed, by following \citep{arakawa20221000}, through the comparison with the correlation matrix represented in Fig.~\ref{fig:MatCorrExp} obtained by calculating the Pearson correlation coefficient\footnote{It is a measure of linear correlation between two sets of data $\{x_i,i=1,...,n\}$ and $\{y_i,i=1,...,n\}$ with $n$ the number of data, defined as $\rho=\frac{ \sum_{i=1}^{n}(x_i-\bar{x})(y_i-\bar{y}) }{\sqrt{\sum_{i=1}^{n}(x_i-\bar{x})^2}\sqrt{\sum_{i=1}^{n}(y_i-\bar{y})^2}}$, where $\bar{x}=\frac{1}{n}\sum_{i=1}^n x_i$ and $\bar{y}=\frac{1}{n}\sum_{i=1}^n y_i$ are the mean values.} for all spider silks.
Observe that  from the database \citep{arakawa20221000} we considered only the data where the searched output and the considered input are reported simultaneously. Thus, since there are some experimental properties missing for some silks of the database, the number of silks composing the training set is different for each considered output. This approach allows us to consider for each target output the maximum number of available information.

%SI OTTIENE SET DI ESPRESSIONI, SELEZIONE ESPRESSIONI
The EPR technique has returned a series of equations (expressions) for each variable searched as output, represented in the figures with increasing numbers on the horizontal axis in Fig.~\ref{fig:two-scales}(d,e,f). These models represent the Pareto front consisting of best formulae considering parsimony (simple expression) and accuracy in a single formulae space exploration. From the whole expression set reported in the Appendix, the most suitable equations (reported in Table~\ref{tab:2scales_expr}) were selected considering not only parsimony and performance, but also analyzing the physical interpretation of the experimental matrix correlations in Fig.~\ref{fig:MatCorrExp}).

%in input first all the Meso scale (see Fig.~\ref{fig:two-scales}(d)) and then all the Micro scale quantities (see Fig.~\ref{fig:two-scales}(g)). In the schemes in Fig.~\ref{fig:two-scales}(a,d,g)  we reported the strategy to obtain each output (box with dotted outline) starting from experimental inputs (box with solid outline) with the arrows indicating each experimental variable at the lower scale used as input to obtain each variable at the upper scale through the EPR method.

\section{Results and Discussion}

%DISCUSSIONE SULLA SATURAZIONE - SPOSTARE

\subsection{Meso from micro}
\begin{figure}	\centering
	\includegraphics[width=1\textwidth]{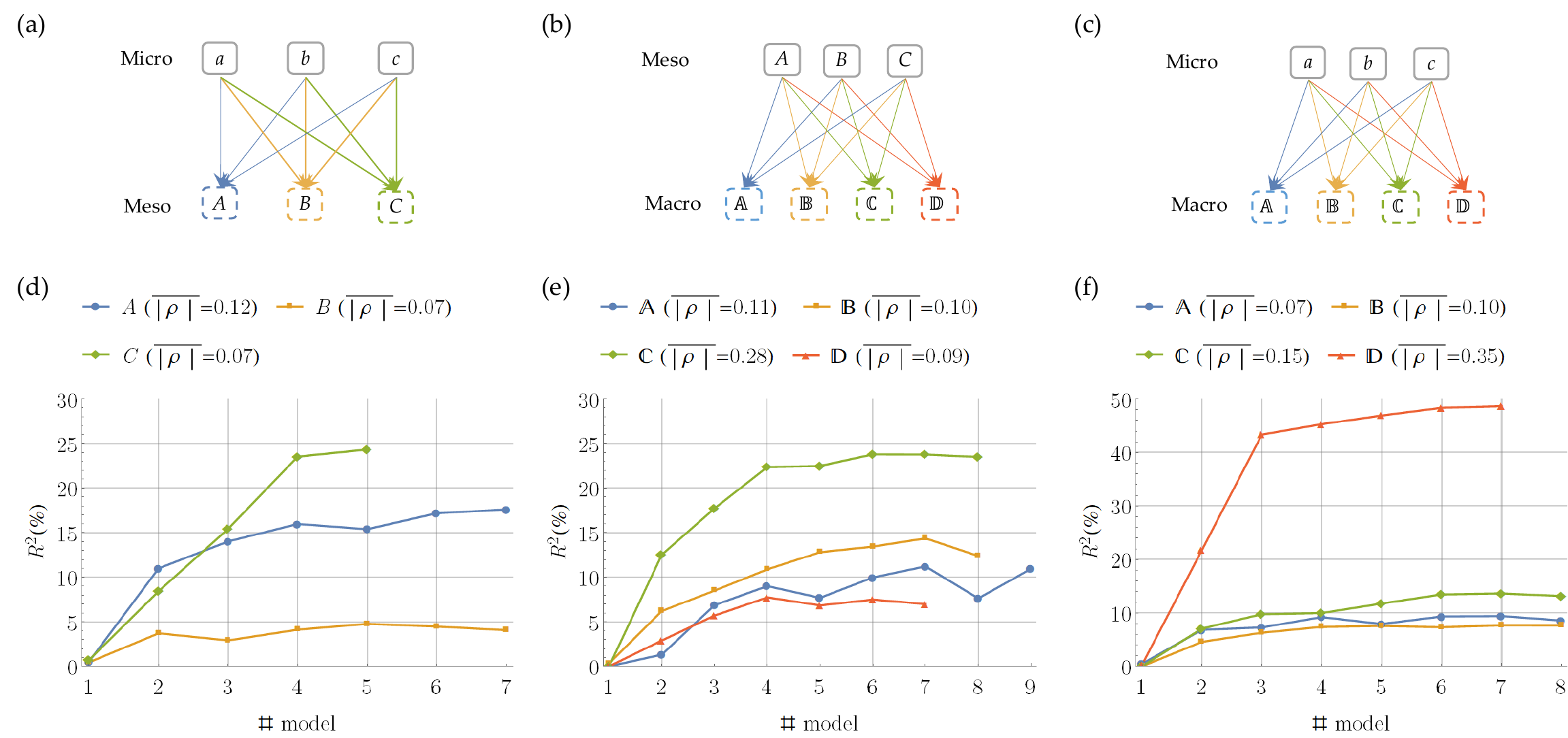}
	\caption{\label{fig:two-scales} 
		Prediction of material properties using two scales at a time:
		(a,d) meso from micro experimental properties, 
		(b,e) macro from meso experimental properties,
		(c,f) macro from micro experimental properties.
		(a,b,c) Scheme of the strategy to obtain each quantity:	solid (Dashed) box indicates experimental (obtained from EPR) quantities.
		(d,e,f) EPR Model performance in terms of $R^2$ plotted against the number of the found model.}
\end{figure}

Firstly, the meso scale properties have been calculated using all the micro scale quantities as independent variables (see Fig.~\ref{fig:two-scales}(a)) according to the recalled equation $ Y=\sum_{j=1}^m  f(x,a_j)+a_0$. 
The results of the accuracy are reported in Fig.~\ref{fig:two-scales}(d) and the resulting functional dependencies are reported in Table \ref{tab:2scales_expr}. 
In Figure \ref{fig:two-scales} we also report the mean of the absolute value of the Pearson correlation $\overline{|\rho|}$ among each dependent variable and the independent variables as a measure of the effective existence of experimental correlation among the considered variables. 
We remark that the EPR algorithm, avoiding overfitting, correctly found functional dependence that exhibits low performances in terms of $R^2$ in the case where also the experimental correlations are low. 

As a general qualitative description of the considered numerical tests (see in particular the variables $A$ and $C$ in Fig.~\ref{fig:two-scales}(d)), we may typically distinguish two regimes of the performance curves. In the first regime, the performance increases rapidly with the number of expressions and thus with the model's complexity. In the second regime, the performance curves stabilize in a saturation band. This indicates an easy way of selecting an optimal model complexity.

Regarding the selected functional dependence, first, we observe that the crystallinity $A$ decreases with $b$, in accordance with the general correlation matrix (Fig.~\ref{fig:MatCorrExp}). The presence of the bias term is coherent with the value of $R^2=11\%$, since, as recalled before, the bias may compensate for the lack of relevant inputs in the model. 

The birefringence $B$ shows a very low accuracy $R^2<5\%$. This can be interpreted by observing that the experimental result 
shows a very Pearson correlation $\overline{|\rho|}=0.07$. We remark that, in this case, the EPR method avoided data overfitting that could have resulted in more performant, but physically misleading expressions. We therefore conclude, in this case, that this meso scale quantity cannot be predicted starting from the considered micro scale properties and we consider instead the $B$ as an independent variable to compute the macro scale quantities in the following.

On the other hand, in the case of the Thermal Degradation Temperature $C$, the EPR found expressions with higher $R^2$. In this case, the selected expression provides a quantitative estimate of the target quantity with a trend increasing with $a$ and $c$, according to the experimental correlation matrix (Fig.~\ref{fig:MatCorrExp}).

\begin{table}[]\centering
	\begin{tabular}{cccc}
		Scale & Expression &$R^2(\%)$&\# model\\
		\hline \\[-.2cm]
		Meso from micro&$A=	3.5562\frac{1}{{b}}+0.10262$& 11 & 2 \\[.2cm]
		&$C=	3186.7046\frac{1}{{a}}+0.86787{a}{c}^{0.5}+45.2672	$ & 23.54& 4 \\[.2cm]
				\hline \\[-.2cm]
			Macro from meso& $\mathbb{A}=0.091301\frac{B^{0.5}}{{A}^{}}+29.2668{A} $ &9.01 & 4  \\[.2cm]
		& $\mathbb{B}=	0.013837\frac{B^{0.5}}{{A}}+0.014276{A}{C}$ & 12.81& 5 \\[.2cm]
		& $\mathbb{C}=	0.81928\frac{{A}^{0.5}{C}}{B}$ &  22.37 & 4 \\[.2cm]
		\hline \\[-.2cm]
		Macro from Micro &
		 $\mathbb{D}=	+0.061926\frac{{b}^{}}{{c}^{}}+0.0047393$ &43.35&3 \\[.2cm]
	\end{tabular} 
	\caption{\label{tab:2scales_expr} Prediction across two scales: selected explicit expressions}\end{table}

\subsection{Macro-meso}

As a second test, the macro properties have been calculated using all the meso scale quantities as independent variables (see Fig.~\ref{fig:two-scales}(b)) according to the usual $
\mathbb{Y}= \sum_{j=1}^m  f(X,a_j)+a_0$. The results of the accuracy are reported in Fig.~\ref{fig:two-scales}(e) and the resulting functional dependencies are reported in Table \ref{tab:2scales_expr}.
 
Regarding the Young's modulus $\mathbb{A}$, the chosen expression correctly reports the monotonic growth with crystallinity $A$, as can be immediately deduced by comparing the derivative of the expression for $A>0$ with the experimental correlation matrix.

Regarding the limit stress $\mathbb{B}$, the selected expression correctly reports the highest experimental correlation, namely the positive one with the Birefringence $B$.

The expression chosen for the diameter $\mathbb{C}$ has the highest accuracy among the macro-meso case ($R^2=22.37\%$) with a very simple expression composed of only a single term that includes all three variables at the meso scale. The correlation is positive for $A$ and $C$ and negative for $B$ in accordance with the experiments. The relatively high performance of the EPR method of this particular case corresponds to fairly relevant experimental correlations $\rho=0.23,-0.39,0.22$ between $\mathbb{C}$ and $A,B,C$, respectively.

Eventually, we consider the selected expression for the supercontraction  $\mathbb{D}$. In this case, we are not able to produce a good estimate of the target output from the meso variables ($R^2<8\%$). This yet is in agreement with the general matrix of experimental correlations with $\overline{|\rho|}=0.07$.

\subsection{Macro-micro}

Eventually, we consider the possibility of direct dependence between macro and micro variables. Thus, the macro properties have been calculated also using all the micro scale quantities as independent variables (see Fig.~\ref{fig:two-scales}(c)) according to equation $\mathbb{Y}= \sum_{j=1}^m  f(x,a_j)+a_0.$ The results of the accuracy are reported in Fig.~\ref{fig:two-scales}(f) and the resulting functional dependencies are reported in Table \ref{tab:2scales_expr}.

%%%%%% macro-micro

In this case, regarding the Young's modulus ($\mathbb{A}$) and the limit stress ($\mathbb{B}$) the values of $R^2$ are generally very low and this is in agreement with the lack of relevant experimental correlations among these quantities. For the diameter ($\mathbb{C}$), the $R^2$ is only slightly higher, again reflecting lightly higher experimental correlations.
On the other hand, the supercontraction $\mathbb{D}$ is predicted with a relatively high accuracy ($R^2>40\%$), and the selected expression ($R^2=43.35\%$) provides a reasonably precise quantitative estimate of the supercontraction. Moreover, the expression is very simple and includes the two most relevant experimental correlations between the supercontraction and the micro scale properties, {\it i. e.} the positive one with $b$ and the negative one with $c$.
 
%Observe that from one side the latter expression let us deduce the functional dependence of the variables across two scale, but from a modelling point of view, the functional dependence across the scale is still unclear. 

%On the other hand it is interesting to focus on the supercontraction properties. Indeed the numerical analysis indicate a direct relation of such Macro quantity on the Micro properties This suggest the possibility of considering in the modelling of the Supercontraction, a direct functional dependence between the macro and the micro properties, without considering the properties of the Meso scale.

By employing this direct macro-micro deduction, from one side we obtain a relatively precise estimation of the supercontraction property that was missing from the macro-meso analysis, but from a modelling point of view, we deduce the possibility of modelling the supercontrction as a macro variable with a direct functional dependence from the micro ones. 
\vince{Moreover, the low accuracy in predicting the other macro variables ($\mathbb{A}$,$\mathbb{B}$,$\mathbb{C}$) directly from the micro ones, enlighten the importance of the meso scale structures in generally determining the macro properties of the material, as expected from the classical hierarchical dependence. For the particular case of the spider silk, this reflects established results in literature pointing out the dependence of the silk thread macroscopic behavior from the secondary structures of the proteins \cite{yazawa2020simultaneous,hayashi1999hypotheses}, here described by the mesoscale variables.
}

\section{Theoretical vs experimental correlations}
%Gli R2 delle 35 silk sono 0.22,0.09,0.003,0.48,0.06,1,0.2,1,1,1.\\

While the objective of this paper is general and mainly related to the exhibited possibilities of obtaining information on the considered physical properties, in this section, we show operatively the possibility of comparing the experimental and theoretical results. In general, we anticipate that the coefficient of determination of expressions found by the EPR method is generally low if compared with other frameworks where EPR was applied \cite{berardi2008development,enriquez2023encapsulating}, but this was expected for the study case of spider silks, as in biological materials a high intrinsic variability for experimental observations is known \cite{cook2014biological}.
Also, for this particular material, a meaningful variability of the mechanical property of silks taken from the same individual under similar conditions is well recognized (see e.g. \cite{madsen1999variability}). Further, the characteristics of the spider silks have high sensitivity to a large number of parameters, among which starvation, reeling speed \citep{madsen1999variability} other than the more expected spider species~\cite{boutry2010evolution}, type of silk (among the up seven different ones that some spiders can produce \cite{vollrath1992spider,gosline1994elastomeric}), environmental conditions \cite{plaza2006thermo} and hydration conditions \cite{agnarsson2009supercontraction}.
In a very recent work \cite{greco2022artificial}, the variability of spider silk properties has been directly compared to that of carbon fibers, and significantly higher variability in spider silk in all properties considered has been reported.
For these reasons, even if the $R^2$ of the expressions found by means EPR is generally not as high as other frameworks, the performance of the data modelling strategies are considered satisfactory. On the other hand, we remark again that the analysis of the effective statistical properties of the theoretical results of spider silks is out of the aim of this paper and is the subject of a forthcoming research of the authors.

Going to the considered case of spider silks, 
we are now in the position of deducing the theoretical meso and macro response based on the only micro properties. According to the previous reasoning, also the meso variable $B$ (birefringence) is considered here as an independent variable. Moreover, coherently with the hierarchical assumption of our model, we first deduce the meso scale variables by the Micro ones and then, based on previous analytical results, we deduce the macro variables. On the other hand, in the special case of the supercontraction $\mathbb{D}$, we assume that it directly depends on micro variables. Notice that all these relevant physical information have been deduced by previous data modelling.

\begin{figure}	\centering
	\includegraphics[width=.8\textwidth]{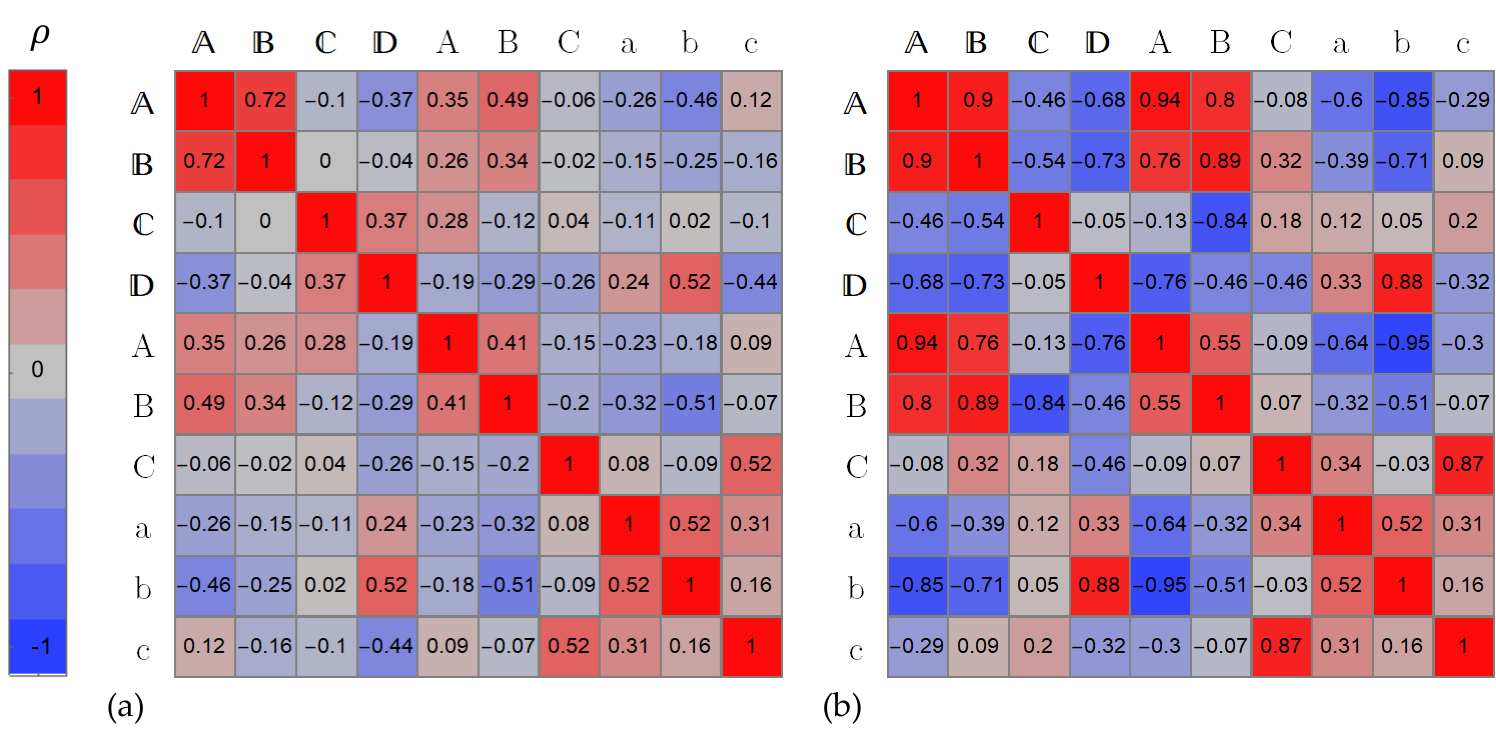}
	\caption{\label{fig:MatCorr} Pearson correlations among the material properties at the three scales. (a)  Experimental correlations obtained considering a subset of silks for which all the analyzed properties are reported simultaneously (b) Correlations among the material properties obtained from the data modelling EPR approach (macro and meso) starting from the known micro experimental properties.}
\end{figure}

Regarding the experimental data, we consider a subset of the silks analyzed by \cite{arakawa20221000} and in particular only those for which all the 10 considered properties (see Table \ref{tab:quantities}) are known simultaneously (the so obtained subset consists of 35 silks). 
%On the other hand, based on the experimental properties of suck silks considered as independent variables we think our analysis, {\it i. e.} the Micro properties and the Birefringence $B$ and by employing the explicit functional dependence reported in Table \ref{tab:2scales_expr} to compute The remaining Meso properties and all the Macro properties, we obtain 35 theoretical silks to compare with the subset of silks where all the properties are obtained by experiments.
As a possible comparison between the theoretical and experimental data sets, we consider the Pearson correlation coefficients for each pair of properties. The results reported in Fig. \ref{fig:MatCorr} show a significant correspondence almost extensible to all the data and a satisfying result in terms of the values of the correlation coefficients.
To get a global comparison we also adopt a positive definite relative error
\begin{equation}
e_r=\frac{|e_a|}{e_m}
\end{equation}
where $e_a=\rho_t-\rho_e$ is the absolute error, $\rho_t$ and  $\rho_e$ are the theoretical and experimental Pearson coefficients, respectively. Here $e_m$ is the mean error that since $\rho_m$ and $\rho_e$ range in the interval $(-1,1)$, we assume as $e_m=1$. The average value of the relative error by considering all the possible pairs of the correlation matrix $\overline{e_r}=0.33$, with $0<\overline{e_r}<2$, indicates that the functional dependence found by the EPR method reproduces in a reasonably accurate way the experimental correlations among the properties of spider silks.

\vince{Eventually, as evidenced in \cite{linka2023new}, an important extension of the proposed approach would be to consider a Bayesian framework for the uncertainty quantification in order to compute each output in terms of statistical distribution with a mean and a confidence interval by also taking into account the input data variability.}

\section{Conclusions}

We showed the possibility of deducing, based on a Genetic Programming approach, data modelling techniques particularly suitable for the deduction of analytical models for multiscale problems.
Our approach is based on the Evolutionary Polynomial Regression (EPR) method, which as we showed lets us deduce models that are both accurate and simple, able to describe the dependence of macro scale variables from the one at lower scales, in their hierarchical order. The best performing models are those located on the Pareto dominance front, which takes into account both accuracy and parsimony and are returned by the EPR algorithm. The final choice of the model can be then based on physical considerations. 

To explicitly show the possibility of acquiring physical insight in a complex multiscale problem, and to evidence the key advantages of our multiscale approach compared to classical, non-physically based techniques, we referred to the materials science field and in particular to the spider silk: a biological material with exceptional properties hugely analyzed also in the spirit of bioinspiration. The choice of this specific case is due to the observation that such remarkable properties are strictly based on an evolutionary hierarchical optimization and the macroscopic spider silk behavior is the result of noteworthy mesostructures emerging from the aggregation of amino acids at the molecular scale. For this intriguing and very complex material optimization case many phenomena are still unclear. We then used this paradigmatic example to show how the presented data modelling approach can be useful in several directions: determine dependent and independent variables, indicate their hierarchical organization, deduce explicit relations among different groups of variables. Furthermore, we showed that the proposed approach let us overcome the overfitting problem typically observed in the analysis of big data within the ANN framework diffusely adopted in this field.  

Based on this, new physical knowledge is acquired, that can be used as a starting point for determining new analytical models, suggesting new experiments, and applying more focused data modelling analysis. In this sense, we assume that Machine Learning or Artificial Intelligence can be impactful for scientific knowledge only if the data modelling approaches are in continuous synergy with the scientific interpretation of the results.
We argue thus that a new mixed GP - theoretical approach can be a new fruitful approach in material science, but also in fields as diverse as biology and medicine.

	\section*{Declaration of Competing Interest}
The authors declare that they have no known competing financial interests or personal relationships that could have appeared to influence the work reported in this paper.	

\section*{Acknowledgments}
Funding: VF and GP have been supported by GNFM (INdAM), GP has been supported by the Italian Ministry MIUR-PRIN project 2017KL4EF3 and by PNRR, National Centre for HPC, Big Data and Quantum Computing (CN00000013) - Spoke 5 ''Environment and Natural Disasters'' and NMP by the European Commission under the FET Open “Boheme” grant no. 863179 and by the Italian Ministry of Education MIUR under the PRIN-20177TTP3S.\\

%\clearpage

%\section*{References}

\bibliography{BibliographyML}

%\printbibliography

\appendix \label{appendix}
\section{EPR Expressions}
%The complete list of the formulae obtained from the EPR tests for each case study WDN is presented in the supplemental material.

		\subsection{Meso-micro}

\subsubsection{Cristallinity}
\begin{subequations}
	\renewcommand{\theequation}{\theparentequation.\arabic{equation}}
	\begin{align}
		A&=	0.19253	\\
		A&=	3.5562\frac{1}{{b}}+0.10262	\\
		A&=	3.9403\frac{1}{{b}}+0.0097339{c}+0.013178	\\
		A&=	0.0010868{b}+2.0441\frac{{c}^{0.5}}{{b}}\\
		A&=	0.007611\frac{{b}}{{a}^{0.5}}+1.961\frac{{c}^{0.5}}{{b}}\\
		A&=	0.0077062\frac{{b}}{{a}^{0.5}}+2.4415\frac{1}{{b}}+0.38037\frac{{c}}{{b}}\\
		A&=	0.0079125\frac{{b}}{{a}^{0.5}}+10.8129\frac{1}{{c}{b}}+0.50353\frac{{c}}{{b}}
\end{align} \end{subequations}

\subsubsection{Birefringence}
\begin{subequations}
	\renewcommand{\theequation}{\theparentequation.\arabic{equation}}
	\begin{align}
		B&=	45.0698	\\
		B&=	88.7638\frac{1}{{b}^{0.5}}+31.2338	\\
		B&=	188.8929\frac{1}{{a}}+54.9397\frac{1}{{b}^{0.5}}+31.4271	\\
		B&=	571.6574\frac{1}{{a}}+1.7581\frac{{a}}{{b}^{0.5}}+19.1995	\\
		B&=	120.3924\frac{{b}^{0.5}}{{a}}+18.2244\frac{{a}}{{b}}+7.3676	\\
		B&=	135.0812\frac{{b}^{0.5}}{{a}}+19.3865\frac{{a}}{{b}}+0.036473{a}+2.329	\\
		B&=	141.865\frac{{b}^{0.5}}{{a}}+17.6973\frac{{a}}{{b}}+0.69564\frac{{a}}{{b}^{0.5}}
\end{align} \end{subequations}

\subsubsection{Thermal degradation temperature}
\begin{subequations}
	\renewcommand{\theequation}{\theparentequation.\arabic{equation}}
	\begin{align}
		C&=	226.8169	\\
		C&=	3.1984{c}+201.1538	\\
		C&=	785.1137\frac{1}{{c}}+13.9926{c}+13.4973	\\
		C&=	3186.7046\frac{1}{{a}}+0.86787{a}{c}^{0.5}+45.2672	\\
		C&=	2482.7659\frac{1}{{a}}+383.6809\frac{1}{{c}}+0.27232{a}{c}+25.2964	
\end{align} \end{subequations}

		\subsection{Macro-meso}

			\subsubsection{Young's Modulus}
\begin{subequations}
\renewcommand{\theequation}{\theparentequation.\arabic{equation}}
 \begin{align}
\mathbb{A}&=	9.4674	\\
\mathbb{A}&=	0.86198B^{0.5}+3.7949	\\
\mathbb{A}&=	0.60871\frac{1}{{A}}+29.105{A}\\
\mathbb{A}&=	0.091301\frac{B^{0.5}}{{A}^{}}+29.2668{A}\\
\mathbb{A}&=	1.3608\frac{B^{0.5}}{{A}{C}^{0.5}}+29.2732{A}\\
\mathbb{A}&=	20.1089\frac{B^{0.5}}{{A}{C}}+1.9687{A}{C}^{0.5}\\
\mathbb{A}&=	2.3655\frac{B}{{A}{C}}+0.098189\frac{1}{{A}}+2.0067{A}{C}^{0.5}\\
\mathbb{A}&=	27.51\frac{1}{{A}{C}}+2.2479\frac{B}{{A}{C}}+2.0041{A}{C}^{0.5}\\
\mathbb{A}&=	10.3522\frac{B^{0.5}}{{A}{C}}+1.3585\frac{B}{{A}{C}}+1.9953{A}{C}^{0.5}
\end{align} \end{subequations}

			\subsubsection{Limit Stress}
\begin{subequations}
\renewcommand{\theequation}{\theparentequation.\arabic{equation}}
 \begin{align}
\mathbb{B}&=	1	\\
\mathbb{B}&=	0.14344B^{0.5}+0.2349	\\
\mathbb{B}&=	0.09374\frac{1}{{A}}+3.1305{A}\\
\mathbb{B}&=	0.013904\frac{B^{0.5}}{{A}}+3.1765{A}\\
\mathbb{B}&=	0.013837\frac{B^{0.5}}{{A}}+0.014276{A}{C}\\
\mathbb{B}&=	0.011517\frac{B^{0.5}}{{A}}+0.0017796{A}{C}B^{0.5}+0.20612	\\
\mathbb{B}&=	0.021735\frac{1}{{A}}+0.0090613\frac{B^{0.5}}{{A}}+0.001894{A}{C}B^{0.5}+0.14047	\\
\mathbb{B}&=	0.012666\frac{B^{0.5}}{{A}}+0.0048689{A}{C}+0.001261{A}{C}B^{0.5}+0.09517	
\end{align} \end{subequations}
			
			\subsubsection{Diameter}
\begin{subequations}
\renewcommand{\theequation}{\theparentequation.\arabic{equation}}
 \begin{align}
\mathbb{C}&=	2	\\
\mathbb{C}&=	81.9474\frac{1}{B}\\
\mathbb{C}&=	177.4203\frac{{A}^{0.5}}{B}+0.051737	\\
\mathbb{C}&=	0.81928\frac{{A}^{0.5}{C}}{B}\\
\mathbb{C}&=	0.00021001{C}+0.80165\frac{{A}^{0.5}{C}}{B}\\
\mathbb{C}&=	0.0037544\frac{{C}}{B^{0.5}}+0.76892\frac{{A}^{0.5}{C}}{B}\\
\mathbb{C}&=	0.011782\frac{{C}}{B}+0.0025068\frac{{C}}{B^{0.5}}+0.76003\frac{{A}^{0.5}{C}}{B}\\
\mathbb{C}&=	0.028813\frac{{C}}{B}+0.69464\frac{{A}^{0.5}{C}}{B}+0.010027\frac{{A}^{0.5}{C}}{B^{0.5}}
\end{align} \end{subequations}
			\subsubsection{Supercontraction}
\begin{subequations}
\renewcommand{\theequation}{\theparentequation.\arabic{equation}}
 \begin{align}
\mathbb{D}&=	0.31695	\\
\mathbb{D}&=	1.4429\frac{1}{B^{0.5}}+0.093513	\\
\mathbb{D}&=	11.8488\frac{{A}^{0.5}}{B}+0.18064	\\
\mathbb{D}&=	0.00085048{C}+10.8458\frac{{A}^{0.5}}{B}\\
\mathbb{D}&=	0.00086571{C}+158.9386\frac{{A}^{0.5}}{{C}^{0.5}B}\\
\mathbb{D}&=	0.00050622{C}+3.8693e-05{C}B^{0.5}+12.6449\frac{{A}^{0.5}}{B}\\
\mathbb{D}&=	0.00067818{C}+2.1548e-05{C}B^{0.5}+172.7934\frac{{A}^{0.5}}{{C}^{0.5}B}
\end{align} \end{subequations}

		\subsection{Macro-Micro}
		\subsubsection{Young's Modulus}
		\begin{subequations}
			\renewcommand{\theequation}{\theparentequation.\arabic{equation}}
			\begin{align}
\mathbb{A}&=	8.5718	\\
\mathbb{A}&=	+150.4503\frac{1}{{b}^{}}+4.8595	\\
\mathbb{A}&=	+103.8759\frac{1}{{a}^{}}+93.8966\frac{1}{{b}^{}}+3.4673	\\
\mathbb{A}&=	+18.4325\frac{{c}^{}}{{a}^{}}+36.55\frac{1}{{c}^{}}	\\
\mathbb{A}&=	+111.3405\frac{{c}^{}}{{a}^{}{b}^{0.5}}+37.8293\frac{1}{{c}^{}}	\\
\mathbb{A}&=	+712.237\frac{{c}^{}}{{a}^{}{b}^{}}+5.6688\frac{{b}^{0.5}}{{c}^{}}	\\
\mathbb{A}&=	+120.4316\frac{{c}^{}}{{a}^{}{b}^{0.5}}+21.7225\frac{1}{{c}^{}}+2.2107\frac{{a}^{0.5}}{{c}^{}}	\\
\mathbb{A}&=	+121.6927\frac{{c}^{}}{{a}^{}{b}^{0.5}}+0.40972\frac{{b}^{}}{{c}^{}}+114.29\frac{{a}^{0.5}}{{c}^{}{b}^{}}
		\end{align} \end{subequations}
		
		\subsubsection{Limit Stress}
		\begin{subequations}
			\renewcommand{\theequation}{\theparentequation.\arabic{equation}}
			\begin{align}
			\mathbb{B}&=	1.2533	\\
			\mathbb{B}&=	+17.8802\frac{1}{{a}^{}}+0.77347	\\
			\mathbb{B}&=	+5.1279\frac{1}{{a}^{0.5}}+3.14\frac{1}{{c}^{}}+0.025086	\\
			\mathbb{B}&=	+146.7895\frac{1}{{a}^{0.5}{b}^{}}+0.014999{b}^{}+0	\\
			\mathbb{B}&=	+0.26238\frac{{b}^{}}{{a}^{0.5}{c}^{0.5}}+24.7876\frac{1}{{b}^{}}+0	\\
			\mathbb{B}&=	+0.29999\frac{{b}^{}}{{a}^{}}+0.056394\frac{{b}^{}}{{c}^{}}+24.7595\frac{1}{{b}^{}}+0	\\
			\mathbb{B}&=	+132.185\frac{1}{{a}^{0.5}{b}^{}}+0.052774\frac{{b}^{}}{{a}^{0.5}}+0.056459\frac{{b}^{}}{{c}^{}}+0.037522	\\
			\mathbb{B}&=	+0.44217\frac{{b}^{}}{{a}^{0.5}{c}^{}}+49.6034\frac{{c}^{0.5}}{{a}^{0.5}{b}^{}}+0.054095\frac{{b}^{}}{{c}^{}}	
		\end{align} \end{subequations}
		
		\subsubsection{Diameter}
		\begin{subequations}
			\renewcommand{\theequation}{\theparentequation.\arabic{equation}}
			\begin{align}
			\mathbb{C}&=	1.5415	\\
			\mathbb{C}&=	+44.5973\frac{1}{{b}^{}}+0.43669	\\
			\mathbb{C}&=	+1030.4476\frac{1}{{a}^{}{b}^{}}+0.82562	\\
			\mathbb{C}&=	+1367.643\frac{1}{{a}^{}{b}^{}}+0.013763{b}^{}	\\
			\mathbb{C}&=	+8702.7135\frac{1}{{a}^{}{c}^{}{b}^{}}+0.01363{b}^{}+0.17322	\\
			\mathbb{C}&=	+9507.752\frac{1}{{a}^{}{c}^{}{b}^{}}+0.0019587{c}^{}{b}^{}	\\
			\mathbb{C}&=	+9491.6515\frac{1}{{a}^{}{c}^{}{b}^{}}+0.001894{c}^{}{b}^{}+7.7482e-05{a}^{}{c}^{}	\\
			\mathbb{C}&=	+9609.6043\frac{1}{{a}^{}{c}^{}{b}^{}}+0.0014633{c}^{}{b}^{}+7.4705e-05{a}^{0.5}{c}^{}{b}^{}
		\end{align} \end{subequations}
		\subsubsection{Supercontraction}
		\begin{subequations}
			\renewcommand{\theequation}{\theparentequation.\arabic{equation}}
			\begin{align}
\mathbb{D}&=	+0.32479	\\
\mathbb{D}&=	+0.0055013{b}^{}+0.097111	\\
\mathbb{D}&=	+0.061926\frac{{b}^{}}{{c}^{}}+0.0047393	\\
\mathbb{D}&=	+0.53648\frac{1}{{c}^{}}+0.050578\frac{{b}^{}}{{c}^{}}	\\
\mathbb{D}&=	+0.8009\frac{1}{{c}^{}}+0.0072816\frac{{a}^{0.5}{b}^{}}{{c}^{}}	\\
\mathbb{D}&=	+15.9524\frac{1}{{c}^{}{b}^{}}+0.008755\frac{{a}^{0.5}{b}^{}}{{c}^{}}	\\
\mathbb{D}&=	+430.4092\frac{1}{{a}^{}{c}^{}{b}^{}}+0.0090635\frac{{a}^{0.5}{b}^{}}{{c}^{}}
	\end{align} \end{subequations}
		
\end{document}